\documentclass{aa}

\usepackage{natbib}
\usepackage{graphicx}
\usepackage[varg]{txfonts}
\date{Received 11 November 2014 / Accepted 12 April 2015}
\usepackage{longtable}
\authorrunning{J.~K.~J{\o}rgensen et al.}
\titlerunning{Molecule sublimation as a tracer of protostellar accretion}

\begin{document}

\title{Molecule sublimation as a tracer of protostellar accretion}\subtitle{Evidence for
  accretion bursts from high angular resolution C$^{18}$O images}

\author{J.~K. J{\o}rgensen\inst{1}, R.~Visser\inst{2},
  J.~P.~Williams\inst{3}, \and E.~A.~Bergin\inst{4}} \institute{Centre
  for Star and Planet Formation, Niels Bohr Institute \& Natural History Museum of
  Denmark, University of
  Copenhagen, {\O}ster Voldgade 5--7, DK-1350 Copenhagen {K}., Denmark
  \and European Southern Observatory,
  Karl-Schwarzschild-Str. 2, D-85748, Garching, Germany \and Institute
  for Astronomy, University of Hawaii at Manoa, Honolulu, HI 96822,
  USA \and Department of Astronomy, University of Michigan, 1085
  S. University Ave, Ann Arbor, MI 48109-1107, USA}

\abstract {The accretion histories of embedded protostars are an
  integral part of descriptions of their physical and chemical
  evolution. In particular, are the accretion rates smoothly declining
  from the earlier toward later stages or in fact characterized by
  variations such as intermittent bursts?} {We aim to characterize the
  impact of possible accretion variations in a sample of embedded
  protostars by measuring the size of the inner regions of their
  envelopes where CO is sublimated and relate those to their
  temperature profiles dictated by their current luminosities.}
{Using observations from the Submillimeter Array we measure the
  extents of the emission from the C$^{18}$O isotopologue toward 16
  deeply embedded protostars. We compare these measurements to the
  predicted extent of the emission given the current
  luminosities of the sources through dust and line radiative transfer
  calculations.  } {Eight out of sixteen sources show more extended
  C$^{18}$O emission than predicted by the models. The modeling shows
  that the likely culprit for these signatures is sublimation due to
  increases in luminosities of the sources by about a factor five or
  more during the recent 10,000~years -- the time it takes for CO to
  freeze-out again on dust grains. For four of those sources the
  increase would have had to have been a factor 10 or more. The
  compact emission seen toward the other half of the sample suggests
  that C$^{18}$O only sublimates when the temperature exceeds 30~K --
  as one would expect if CO is mixed with H$_2$O in the grain
  ice-mantles.}  {The small-number statistics from this survey
  suggest that protostars undergo significant bursts about once every
  20,000~years. This also illustrates the importance of taking the
  physical evolutionary histories into account for descriptions of the
  chemical structures of embedded protostars.}

\offprints{Jes K.\,J{\o}rgensen, \email{jeskj@nbi.ku.dk}}
\maketitle

\section{Introduction}\label{introduction}
In their earliest stages, solar-type protostars are characterized by
large amounts of cold gas and dust surrounding them. As the young star
begins to heat up and accretion proceeds, this envelope is dissipated
revealing the pre-main sequence star surrounded by a circumstellar
disk. Studies of young stars during the embedded stages are of prime
importance for our understanding of their formation and early physical
and chemical evolution: this is the phase when the star accretes the
bulk of its mass and a circumstellar disk is formed. The physical and
chemical structure of this disk possibly sets the initial conditions
for subsequent planet formation. One of the important questions for
our theories of formation of stars is how accretion proceeds: is it
taking place at a relatively constant rate throughout the protostellar
evolution or strongly varying, e.g., related to the formation of the
disk and possible fragmentation within it? Indirect evidence suggests
that the accretion rates may be varying through the protostars' early
years -- for example observations of outflows that in some cases show
clear knots or bullets that can be attributed to changes in the
underlying accretion rates \citep[e.g.,][]{reipurth89,arce13}. This
paper presents maps of the C$^{18}$O emission on scales of a few
hundred to 1000 AU toward the dense envelopes of a sample of 16
embedded protostars with the Submillimeter Array (SMA). The aim of
this work is to address whether possible variations in the
protostellar accretion rates can be traced by their chemical
signatures.

Strong luminosity bursts have for some time been recognized as one of
the characteristics of young stars -- at least during specific times
of their pre-main sequence phase where they may appear as FU Orionis
objects at visible wavelengths \citep[see, e.g.,][for a recent
review]{audardppvi}. Such outbursts can be tied to the accretion onto
and through the central protostellar disks -- and possible
instabilities in those
\citep[e.g.,][]{bell94,armitage01,vorobyov05,zhu09,martin11}. Due to
the significant amounts of obscuring gas and dust, it is still unclear
what happens in the earlier protostellar stages. Photometric
comparisons of nearby star-forming regions at different epochs show
small-scale luminosity variations on time-scales ranging from days to
years \citep[e.g.,][]{rebull14} -- and are able to identify individual
sources undergoing bursts in intervals of 5000--50,000~yr
\citep{scholz13}.  If these bursts are an indication that the
accretion rates are indeed also strongly varying, it would definitely
be a critical aspect of the protostellar evolution.

In the earliest stages the luminosity of the protostar is thought to
be dominated by the release of gravitational energy due to the
accretion onto the central star ($L \sim G M_\ast
\dot{M}/r_\ast$). Generally, protostars are found to show a very broad
luminosity distribution and be under-luminous compared to the
expectation from constant infall toward the center of their natal
cores (see, e.g., \citealt{kenyon90} and \citealt{evans09}; as well as
\citealt{dunhamppvi} for a recent review). This is another piece of
evidence that has been taken in favor of episodic accretion during the
earliest protostellar phases. In that picture, protostars spend most
of their lifetimes in a low accretion rate mode, assembling material
in the disk, and thus appear less luminous than what should be
expected based on accretion from their envelopes directly onto the
central star. They could then accrete significant mass in relatively
short bursts accompanied by increases in their luminosities that
statistically would be relatively difficult to observe. Still, because
of the complex environments of young stars and possible continued
infall onto and through the disk also during the more evolved stages,
this picture may be too simplistic and the luminosity distribution
itself not offering a good constraint \citep[e.g.][]{padoan14}.

Besides the obvious important dynamical consequences of the exact
protostellar accretion histories, they may also have important
consequences on other aspects of the early evolution of protostars --
for example, the chemical processes leading to the formation of larger
molecules. This process involves a complex balance between molecules
freezing out onto dust grains, grain-surface chemistry and eventual
evaporation of the species before they may be (re-)incorporated into
ices in the circumstellar disks \citep[see, e.g.,][for recent
reviews]{herbst09,caselli12}. Variations in the accretion rates and
consequently luminosities of the protostars will affect how long given
molecules are present in solid or gaseous form and what the
time-scales are for the regulating chemical processes at different
points during the protostellar evolution.

An interesting question is how these two issues, the accretion
histories of protostars and corresponding luminosity evolution on the
one hand and chemical evolution on the other, correspond. The strong
dependence of the protostellar luminosity on the accretion rate will
reflect directly in the thermal structure of the envelopes through the
(almost instantaneous) heating of the dust \citep[][]{johnstone13}. As
these times are also characterized by freeze-out and sublimation of
molecules that are strongly temperature dependent, clues may be found
in the composition of ices as well as distributions of molecules in
the gas- and solid-phase
\citep[e.g.,][]{lee07,poteet13,visser12,visser15}. One example of this
is observations that show significant amounts of pure CO$_2$ ices
present around low to moderate luminosity protostars
\citep{kim12}. The presence of pure CO$_2$ ice is a good indication
that significant thermal processing of the ices has taken place prior
to the current evolutionary stage of the protostars. Still, such
observations of the total column density of a specific ice species
only reveal the integrated history of the source -- and cannot be used
as constraints on whether this distillation has taken place during the
protostellar evolution or perhaps prior to the onset of collapse.

Another way of addressing this issue may be through resolved
observations of the distributions of molecules in the
environments. One possible example of this are ALMA observations of
the deeply embedded protostar IRAS~15398--3359
\citep{jorgensen13}. The resolved images show a depression in the
emission of H$^{13}$CO$^+$ in the inner 150--200~AU of the central
protostellar core. One explanation for this depression is that HCO$^+$
is destroyed by reactions with extended water vapor. However, the
HCO$^+$ depression is seen on the scales where the temperature is as
low as 30~K given the current luminosity of the protostar -- well
below the temperature of 90--100~K needed for water to sublimate. A
possibility is that the source has undergone a burst in accretion
during the last 100--1000~years, increasing the luminosity by up to a
factor 100 above what it currently is, causing water to sublimate on
much larger scales. Ideally, one would use similar observations for
many sources to constrain the prominence of such possible bursts: if
such variations are indeed due to accretion variations, the
distributions of even simpler species such as the optically thin
isotopologues of CO may in fact already hold that information.

In this paper we present an analysis of C$^{18}$O emission for a
sample of 16 embedded protostars from previously published/archival
Submillimeter Array observations with the aim of revealing their CO
emission distributions. The paper is laid out as follows:
\S\ref{observations} describes the sample of sources and presents an
overview of the observed images and spectra. \S\ref{analysis} presents
an analysis of the observed maps in contexts of detailed dust and line
radiative transfer models -- with a specific case study of the Class~0
protostar IRAS~03282+3035 setting up a more general analysis for the
sample of 16~sources. The key result, that half of these protostars in
fact show more extended CO emission than what one should expect based
on their current luminosities, is discussed in \S\ref{discussion} --
along with the statistical implications for the frequency of bursts
and the properties of the CO ices. \S\ref{summary} summarizes the main
findings of the paper.

\section{Data}\label{observations}
As basis for this analysis we constructed a sample of embedded
protostars observed using the Submillimeter Array (SMA;
\citealt{ho04}). One of the preferred spectral setups for such
observations at 230~GHz (1.3~mm) is a setting covering the $J=2-1$
transitions of $^{12}$CO, $^{13}$CO and C$^{18}$O. For the purpose of
this paper we focus on C$^{18}$O, which typically shows centrally
concentrated emission toward the location of the protostars. We
utilized observations in the SMA's compact-North and compact
configurations that typically result in an angular resolution of
2--3\arcsec\ at these frequencies. Due to the interferometer's lack of
short-spacings, the observations are typically not sensitive to
emission extended over scales larger than about 15\arcsec.

The sample was put together utilizing the PROSAC study of Class~0
protostars \citep{prosacpaper} as well as the sources from the SMA
archive that were also used as a basis for the analysis of binarity by
\cite{chen13}. This sample comprises predominantly Class~0
  objects -- with a few young Class I objects (i.e., sources with
  $T_{\rm bol} \lesssim 150$~K -- or envelope masses of a few$\times
  0.1~M_\odot$ to a few $M_\odot$). We exclude sources with low
signal-to-noise detections either due to their intrinsic C$^{18}$O
2--1 line strengths or poorer data quality.

A few sources for which observations have not previously been
presented were added. Those include IRAS~03256-3055 observed on
2011/09/24 as part of a larger program to survey the population of
young stars in the NGC~1333 cluster (PI: J.~Williams) as well as
observations of two intermediate mass protostars in Orion, MMS6 and
MMS9 \citep{johnstone03,isrfart}, from observations on 2007/02/07 (PI:
J.~J{\o}rgensen). The full sample of sources is summarized in
Table~\ref{sample}.
\begin{table*}
\caption{Sample of sources}\label{sample}
\begin{center}\begin{tabular}{lllllll}\hline\hline
Source    & $\alpha_{\rm J2000}$ & $\delta_{\rm J2000}$
& $L_{\rm cur}$\tablefootmark{a} & Distance & SMA
reference\tablefootmark{b} & Luminosity reference\\
   & (hh mm ss.s) & (dd mm ss) & ($L_\odot$) & (pc) & & \\ \hline
L1448I2   & 03 25 22.4 & $+$30 45 13 & 1.7  & 235 & \cite{chen13} & \cite{evans09} \\
L1448I3   & 03 25 36.3 & $+$30 45 15 & 8.8  & 235 & \cite{chen13} & \cite{evans09} \\
L1448C    & 03 25 38.9 & $+$30 44 05 & 8.4  & 235 & \cite{prosacpaper} & \cite{green13} \\
IRAS03256 & 03 28 43.4 & $+$31 17 37 & 1.7  & 235 & 11/09/24 (J. Williams) & \cite{evans09} \\
IRAS2A    & 03 28 55.6 & $+$31 14 37 & 20   & 235 & \cite{prosacpaper} & \cite{prosacpaper} \\
SVS13A    & 03 29 03.7 & $+$31 16 07 & 34   & 235 & \cite{chen13} & \cite{evans09} \\
IRAS4A    & 03 29 10.5 & $+$31 13 32 & 9.9  & 235 & \cite{prosacpaper} & \cite{karska13} \\
IRAS4B    & 03 29 12.0 & $+$31 13 08 & 4.4  & 235 & \cite{prosacpaper} & \cite{karska13} \\
IRAS03282 & 03 31 20.9 & $+$30 45 30 & 1.1  & 235 & \cite{chen13} & \cite{evans09} \\
TMR1      & 04 39 13.9 & $+$25 53 21 & 3.8  & 140 & 10/11/29 (H.-W. Yen) & \cite{karska13} \\
L1527     & 04 39 53.9 & $+$26 03 10 & 1.9  & 140 & \cite{prosacpaper}  & \cite{karska13} \\
MMS6      & 05 35 26.0 & $-$05 05 43 & 27   & 420 & 07/02/24 (J. J{\o}rgensen) & (see below) \\
MMS9      & 05 35 23.4 & $-$05 01 31 & 6.3  & 420 & 07/02/24 (J. J{\o}rgensen) & (see below) \\
IRAS15398 & 15 43 02.2 & $-$34 09 07 & 1.6  & 155 & \cite{chen13} & \cite{karska13} \\
CrA32     & 19 02 58.7 & $-$37 07 36 & 1.3  & 130 & \cite{peterson11} & \cite{peterson11} \\
B335      & 19 37 00.9 & $+$07 34 10 & 1.3  & 100 & \cite{prosacpaper} & \cite{launhardt13} \\  \hline
\end{tabular}\end{center}

\tablefoot{\tablefoottext{a}{Bolometric luminosities compiled from recent Herschel and Spitzer
surveys with references in the final column. For the
two sources in Orion, MMS6 and MMS9, some differences are found in
literature. We adopt the value of 27~$L_\odot$ for MMS6 from \cite{manoj13} and derive the internal luminosity
of MMS9 by scaling by the ratio of the 70~$\mu$m fluxes of the two
sources reported by
\cite{billot12} (see also discussion of the relation between 70~$\mu$m flux and
internal luminosity in \citealt{dunham08}).}
\tablefoottext{b}{Published paper first describing data -- or, if no previous paper, date of observations and PI.}}
\end{table*}

Both the archival data and the new observations were re-reduced using
the MIR package \citep{qimir} and imaged and cleaned using Miriad
\citep{sault95}. In the reduction we followed the standard procedures,
calibrating the complex gains through observations of nearby quasars
(typically two per track), passband calibration through observations
of strong quasars before and/or after the observations and flux
calibrations utilizing planet observations. Continuum sensitivities
(RMS) in 2~GHz bandwidth for these observations were
0.5--1~mJy. Typically the C$^{18}$O 2--1 observations had been
performed by allocating 128--1024 channels to one of the (then)
twenty-four 104~MHz chunks of the SMA correlator placed to cover the
C$^{18}$O transition. This resulted in spectral resolutions of
0.14--1.1~km~s$^{-1}$ with a typical line RMS noise of
30--60~mJy~beam$^{-1}$ averaged over 1~km~s$^{-1}$ channels.

\subsection{Overview of data}
\begin{figure*}
\sidecaption
\includegraphics[width=12cm]{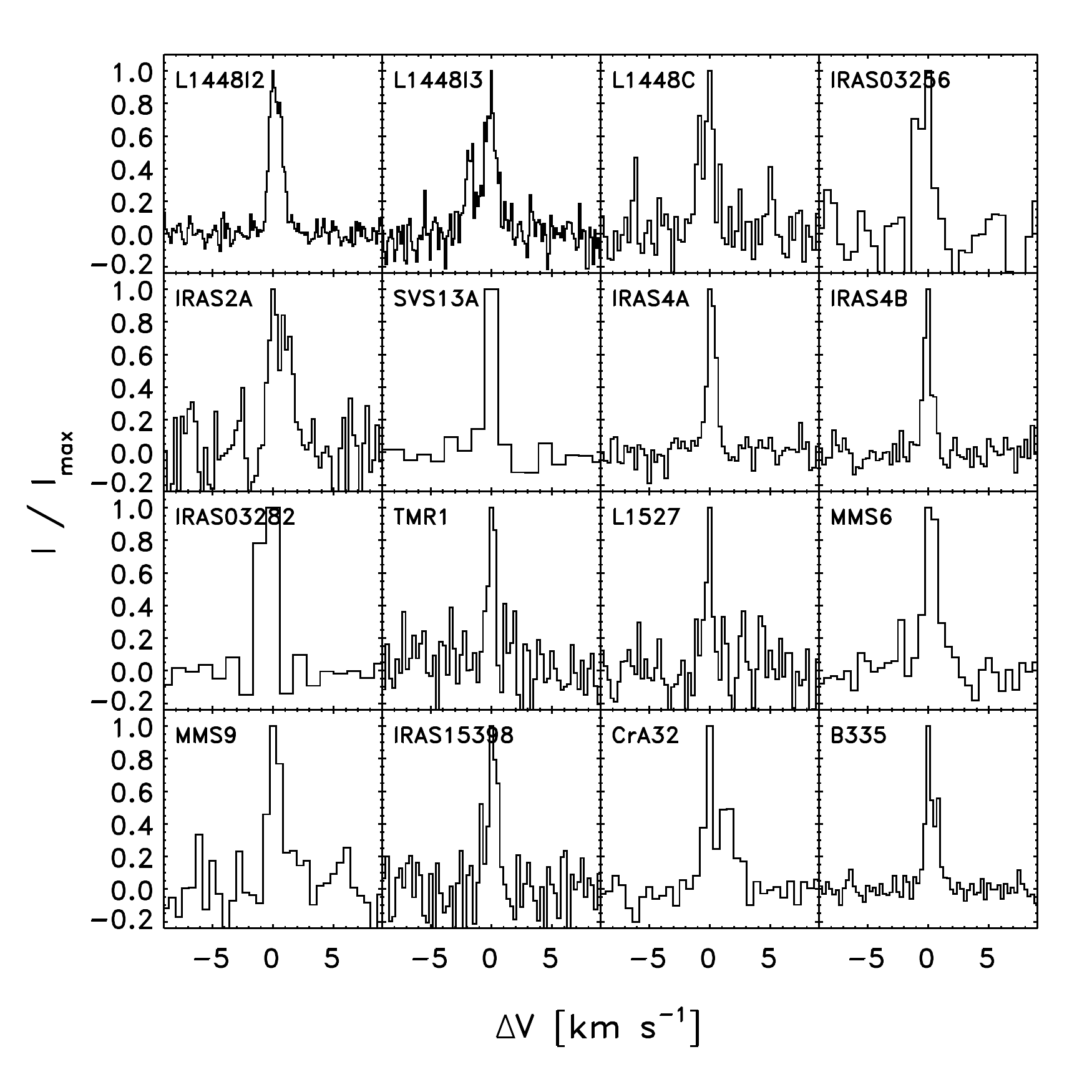}
\caption{C$^{18}$O 2--1 spectra toward the peak of its integrated
  emission for each of the protostars in the
  sample.}\label{c18ospectra}
\end{figure*}

\begin{figure*}
\sidecaption
\includegraphics[width=12cm]{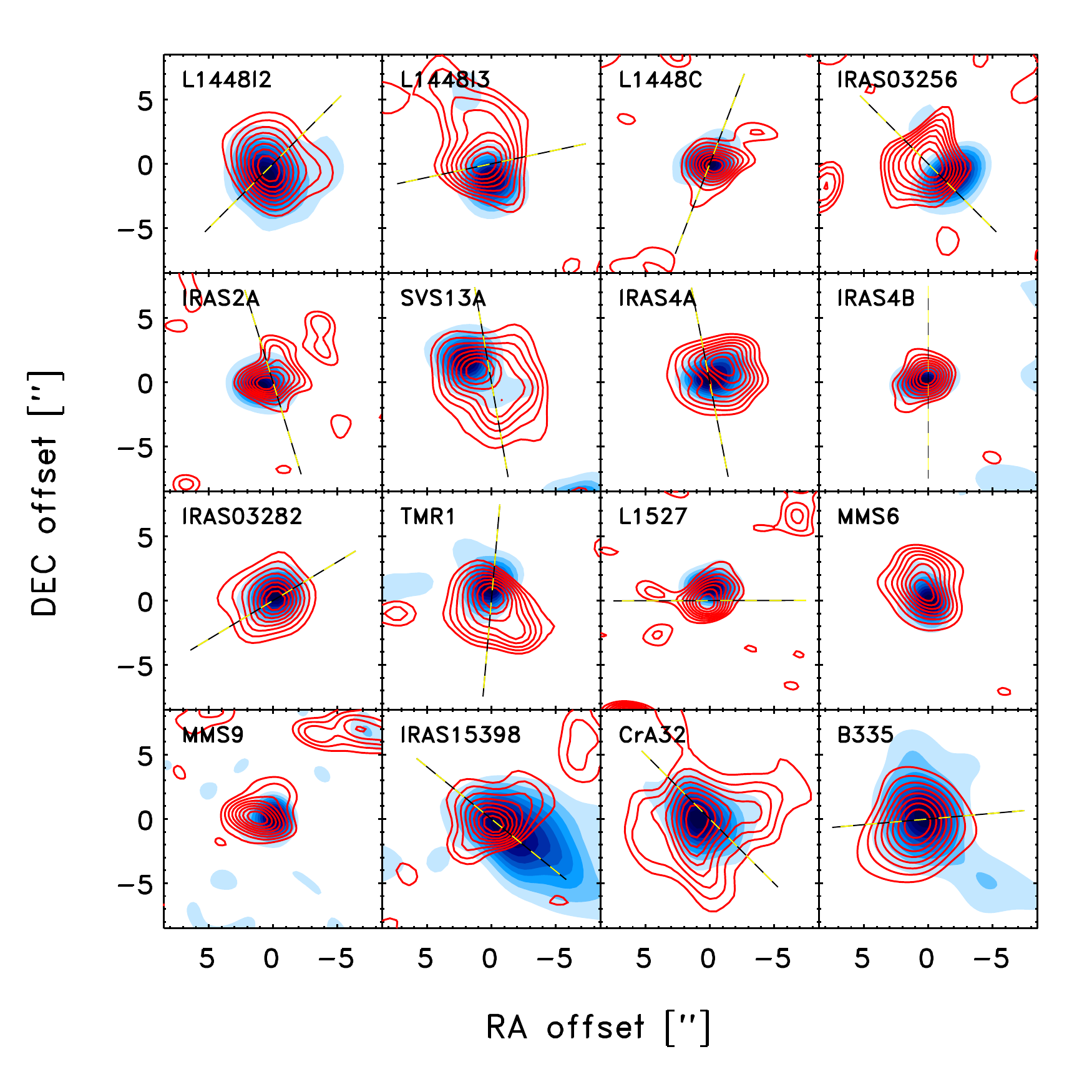}
\caption{Integrated C$^{18}$O 2--1 emission for each of the protostars
  shown as contours on top of continuum maps of each source. The
  contours are shown at steps corresponding to 20\%, 30\%, 40\%
  $\ldots$ of the peak emission in the integrated line maps. The
    dashed lines indicate the outflow axis from SMA $^{12}$CO 2--1
    maps for the sources where this line was observed simultaneously
    with C$^{18}$O 2--1.}\label{c18omaps}
\end{figure*}

Figs.~\ref{c18ospectra} and \ref{c18omaps} present overviews of the
observed C$^{18}$O 2--1 spectra and maps. The spectra were extracted
in the beam toward the peak of the continuum emission for each
source. The maps were made by integrating the emission over $\pm
  1.5$~kms$^{-1}$ around the systemic velocity for each
line.

The spectra show fairly regular profiles with little evidence for
wings or other asymmetries as often is the case for the more common
isotopologues -- as well as in larger scale single-dish maps. For the
maps the emission is also relatively centrally condensed with low
surface brightness extended emission in a few cases -- but
interestingly not for the most prominent outflow sources. Fig.~\ref{c18omaps} also shows that for the sources with some
  structure in the C$^{18}$O 2--1 emission there is no clear
  correlation with the outflow axis probed by $^{12}$CO 2--1.  These
observations likely reflect that the optically thin \emph{integrated}
C$^{18}$O 2--1 emission is relatively little affected by the outflows
present on larger scales but rather associated with the denser parts
of the envelopes on scales $\lesssim 1000$~AU -- in agreement with
what has been inferred from previous high angular resolution
millimeter wavelength studies \citep[e.g.,][]{arce06}. In a few cases
(e.g., IRAS03256 and IRAS15398) small offsets are seen between the
continuum and line emission in Fig.~\ref{c18omaps}: for those sources
it is seen that the line emission is generally better aligned with
location of the central heating source as traced by shorter wavelength
(e.g. Spitzer data) and the SMA continuum emission is more strongly
affected by the dust density and temperature distribution on the
scales of the envelope--disk transition.

In the following we focus on reproducing the extent of the integrated
emission from the C$^{18}$O 2--1 transition -- but further restrict
the analysis beyond what is shown in Figs.~\ref{c18ospectra} and
\ref{c18omaps} and do not include the most extended (low surface
brightness) emission as justified below.

\section{Analysis}\label{analysis}
The main aspect of the analysis of the C$^{18}$O 2--1 emission is to
test whether its extent can be reproduced through simple radiative
transfer models of the envelope structure and protostellar
luminosities. We first present an in-depth analysis of a single source
as a test case (\S\ref{testcase}) before considering the entire sample
of sources in a general manner (\S\ref{approach}).

\subsection{Test case: IRAS~03282+3035}\label{testcase}
As a test case we focus on the IRAS~03282+3035 Class~0 protostar
(IRAS03282 in the following): IRAS03282 is a deeply embedded protostar
in the Perseus molecular cloud driving a northwest/southeast oriented
outflow seen in high resolution millimeter interferometric maps
\citep[e.g.,][]{arce06} and in mid-infrared images from the Spitzer
Space Telescope \citep[e.g.,][]{perspitz}. Its compact C$^{18}$O 2--1
emission is particularly bright and regular, making it a good test
case for the method.

To model the C$^{18}$O 2--1 profile we construct a model for the
envelope based on its submillimeter continuum and line emission --
similar to the approach adopted in, e.g.,
\cite{jorgensen02,evolpaper}. We adopt a 1-dimensional power-law
density profile for the envelope, $\rho \propto r^{-p}$, and calculate
its temperature profile self-consistently based on the luminosity of
the source. For the calculations we utilize the radiation transfer
code \emph{Transphere}\footnote{\tt http://www.ita.uni-heidelberg.de/
  $\sim$dullemond/software/transphere/index.shtml} that solves the 1D
dust radiative transfer problem for absorption and re-emission
following the methods described in \cite{dullemond02}. The
normalization of the envelope density profile is fixed by comparison
to single-dish submillimeter continuum observations (for further
discussion see, e.g., \citealt{evolpaper}). In the following we test
models that fix $p$ at 1.5 (the expectation for a free-falling
envelope) and 2.0 (corresponding to a static isothermal sphere). From
radiative transfer models of submillimeter continuum brightness
profiles from JCMT/SCUBA observations, $p$ is typically found to vary
between these values with uncertainties of $\pm 0.2$ in the
measurements for individual sources
\citep[e.g.,][]{jorgensen02,shirley02,kristensen12}.

\begin{figure*}
\resizebox{\hsize}{!}{\includegraphics{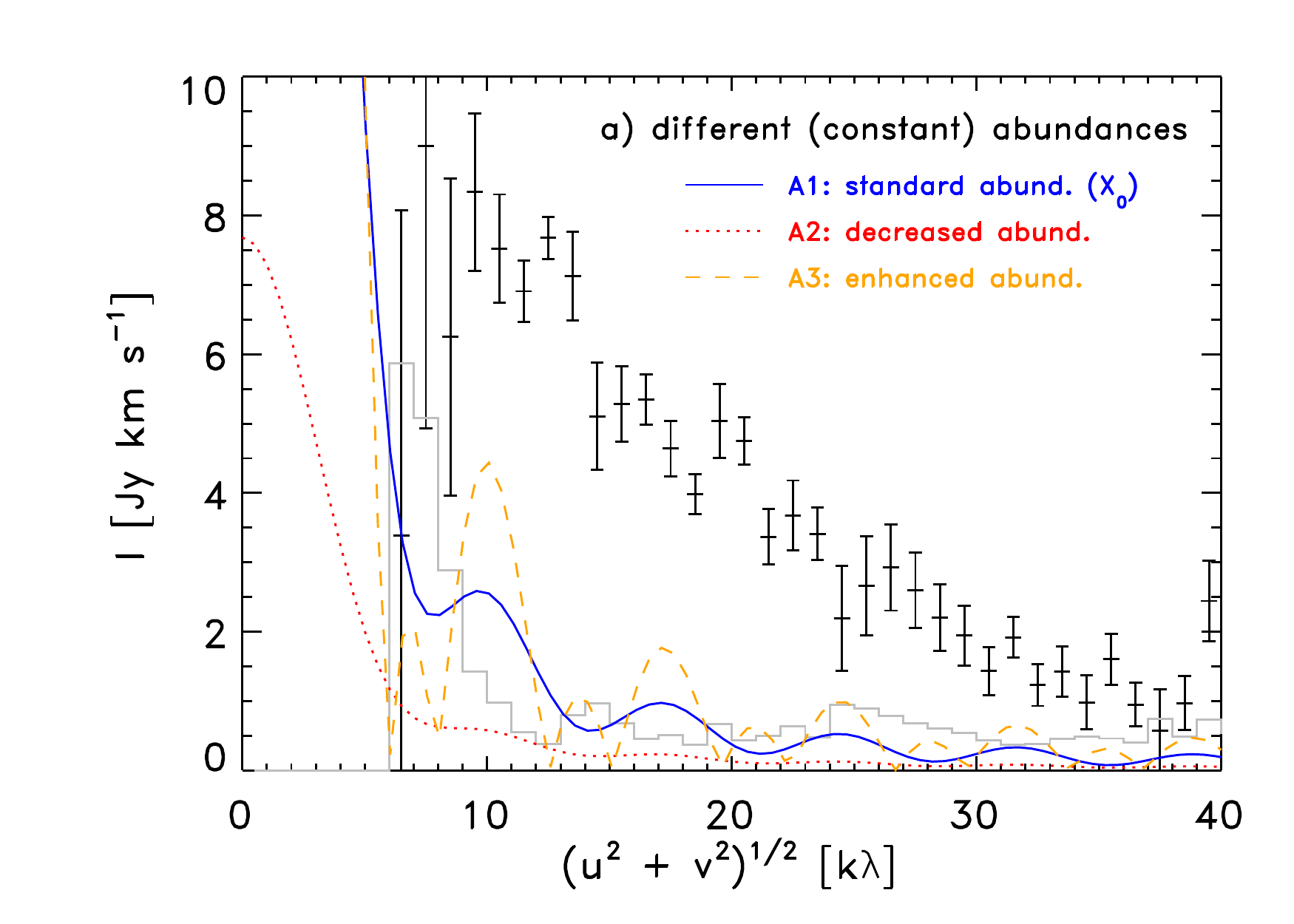}\includegraphics{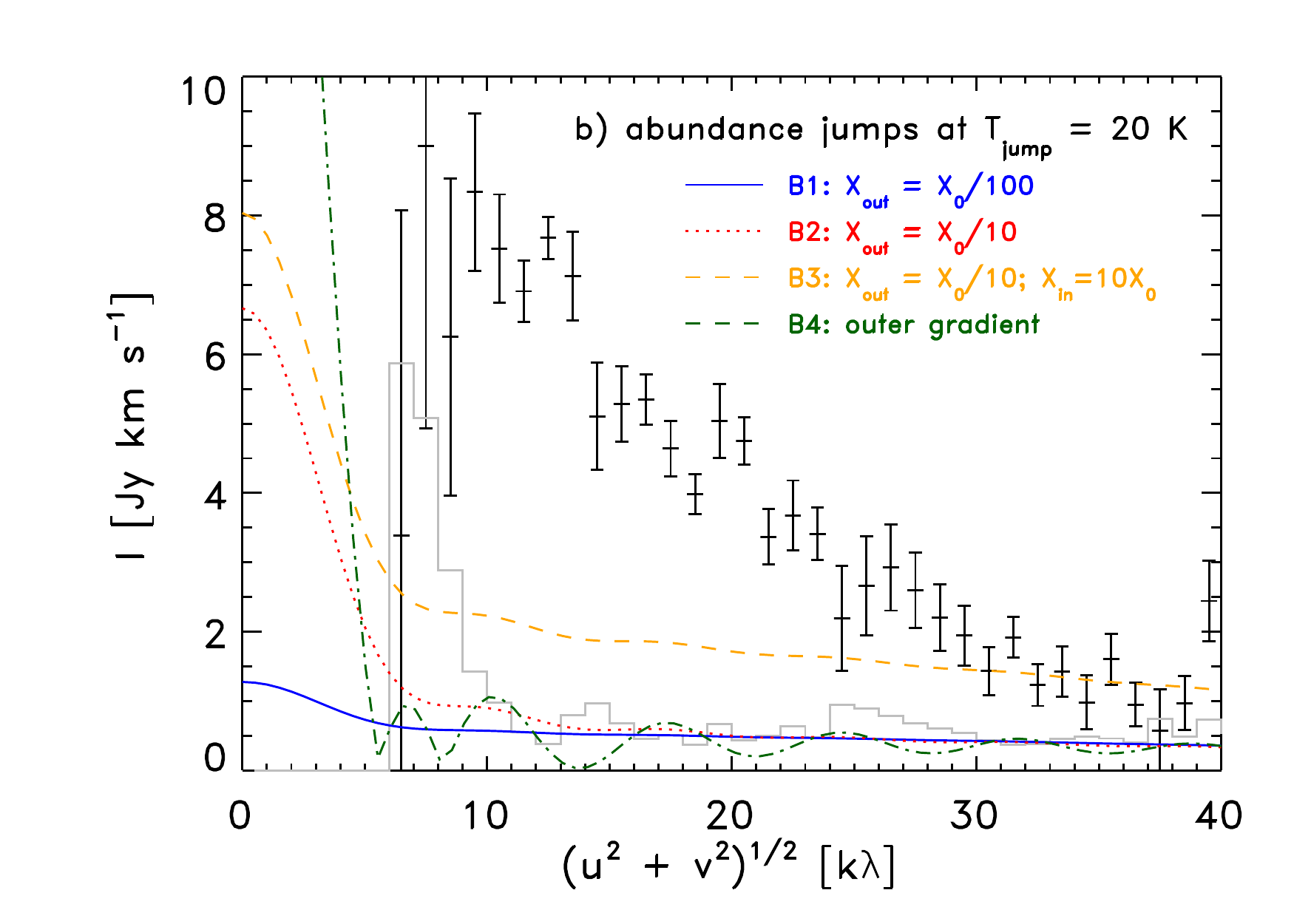}}
\resizebox{\hsize}{!}{\includegraphics{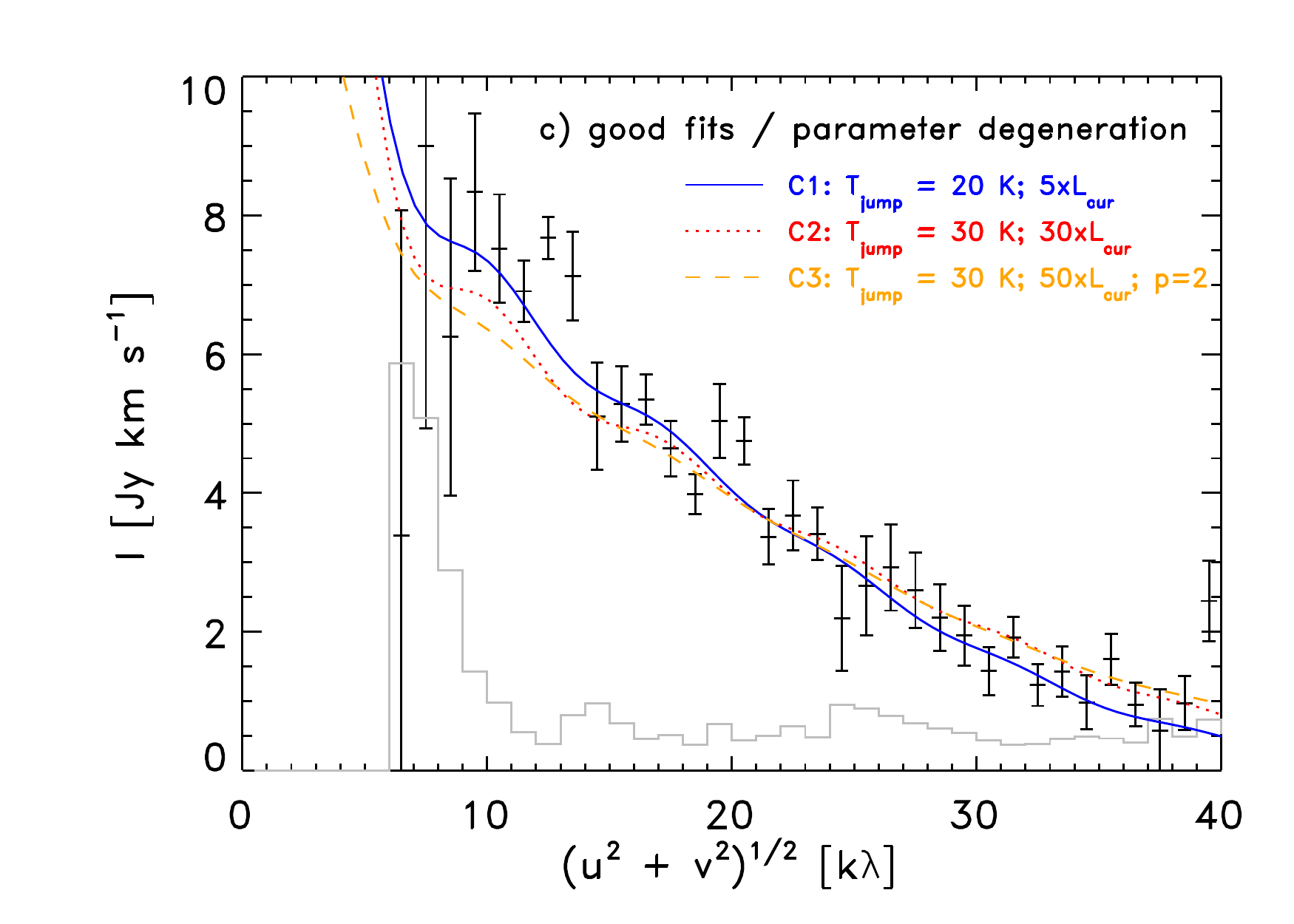}\includegraphics{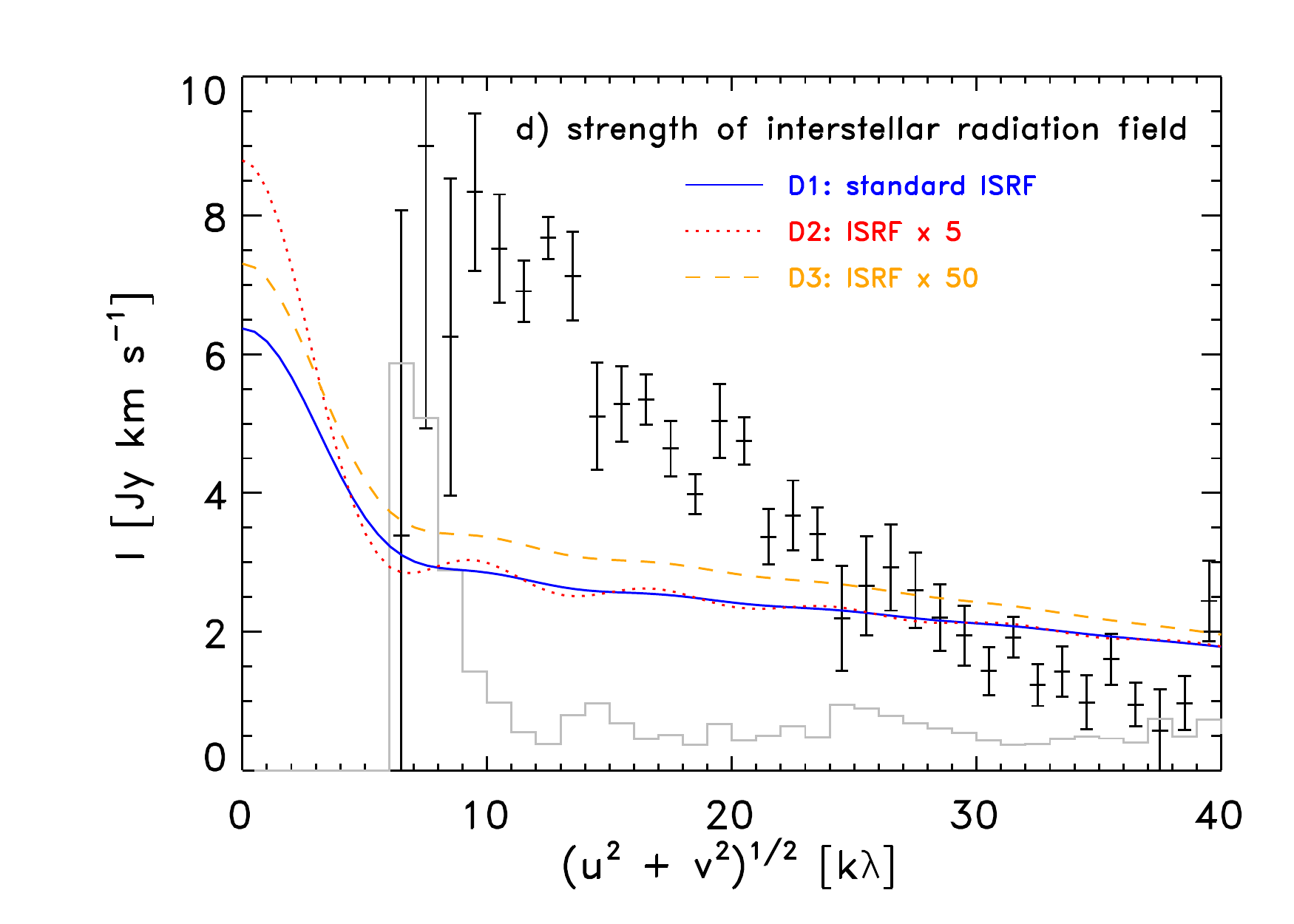}}
\caption{Comparison between different parameterisations of the
  envelope physical and chemical structure (for details of the
  parameters for each model, see Table~\ref{modelsummary}) comparing
  the observed and modeled visibility amplitude as function of
  projected baseline length for IRAS03282. In each panel the observed
  integrated C$^{18}$O emission is seen as datapoints with models as
  lines. The grey histogram indicates the expected amplitude for zero
  signal. In panel \emph{a)} models with constant abundances and a
  luminosity corresponding to the current source luminosity
  (1.1~$L_\odot$). In panel \emph{b)} models with a changing abundance
  at the radius where the temperature increases above 20~K. In panel
  \emph{c)} the source luminosity is increased and CO consequently
  sublimates at larger radii providing the variation in intensity as
  function of projected baselines required to fit the data. Finally
  panel \emph{d)} tests the importance of varying the strength of the
  external interstellar radiation field -- which again is found to
  only affect the outer parts of the
  envelopes.}\label{parameterexploration}
\end{figure*}

The resulting density and temperature profiles from the dust radiative
transfer calculations are then used for line radiative transfer
calculations using the Ratran code
\citep{hogerheijde00vandertak}. Adding the velocity field and an
abundance profile for a given molecule, Ratran solves the full non-LTE
radiation problem to estimate its level populations as function of
location and subsequently ray-traces the solution to calculate
synthetic images that can be compared directly to the
observations. For the CO abundance profile we adopt a simple step
function \citep[e.g.][]{coevollet}. We multiply the synthesized images
with the interferometric primary beam and Fourier transform the images
for the direct comparison to the observations in the
$(u,v)$-plane. Fig.~\ref{parameterexploration} explores the influence
of different models on the CO emission profiles. In the panels of this
figure the data and models are represented by the amplitude of the
integrated emission as function of projected baseline length.

With these choices the model has a number of free parameters: the
envelope inner and outer radius, its density normalization and the CO
abundance profile as function of radius. By restricting the comparison
to specific baselines the inner and outer envelope radii are not
important for the results and we adopt 25~AU and 8000~AU as in
\cite{evolpaper}. The normalization of the density profile and the
absolute abundances are naturally degenerate -- in the sense that a
denser/more massive envelope can fit the data with a lower
abundance. The shape of the abundance profile on these scales is
critical, however.

Fig.~\ref{parameterexploration} compares the observed C$^{18}$O
visibility amplitudes to the predictions from a range of models
summarized in
Table~\ref{modelsummary}. Fig.~\ref{parameterexploration}a illustrates
the simplest models where the C$^{18}$O abundance is kept constant
throughout the envelope at $2\times 10^{-7}$, corresponding to a
canonical CO abundance of $10^{-4}$ -- as well as models where the
abundance is increased and decreased by a factor 8. Two important
conclusions can immediately be drawn from this plot: if the C$^{18}$O
abundance is constant at the canonical level the emission of the 2--1
transition becomes optically thick throughout the envelope -- so that
further increases in abundance do not change the brightness profiles
significantly. For low abundances, the model underproduces the
emission on large scales (short baselines) as well as in single-dish
observations. For all three models little emission is seen on long
baselines compared to the observations that show significant emission
out to baselines of approximately 35--40~$k\lambda$. This indicates
that the abundance must change on small scales -- rather than staying
constant throughout the envelope -- making the emission maps appear
more centrally condensed.

Models with an abundance enhancement of CO by up to two orders of
magnitude (Fig.~\ref{parameterexploration}b) at the radius where the
temperature increases to either 20 or 30~K (corresponding to the CO
sublimation temperature of pure or slightly mixed CO ice) do a
slightly better job in the sense of producing some emission at the
smaller scales/longer baselines -- but are still not able to capture
the full emission profiles. Also, for a canonical CO abundance on
small scales the model severely underproduces the emission on larger
scales, so the change has to occur on larger scales than what would be
predicted given the current luminosities. The figure also
  includes a model with an abundance decreasing through the
  protostellar envelope from the outer edge towards the centre until
  the point where the the temperature reaches 20~K and then increases
  abruptly back to the canonical value. Such models could capture,
  e.g., the gradual freeze-out with increasing densities
  \citep{coevollet,fuente12} inferred on the basis of single-dish and
  (lower resolution) interferometric data. In addition, this model
  does not affect the emission on the longer baselines. Thus,
  the key point from this analysis is that the shape of the visibility
  amplitude curve as function of projected baselines is not affected
  by the exact abundance structure on larger scales. The task is
  therefore to determine what luminosity is required to shift the CO
  sublimation temperature out to the radius where the enhancement is
  seen in the CO emission maps.

Fig.~\ref{parameterexploration}c demonstrates three models that fit
the data well: these represent models where the CO sublimates with an
abundance enhancement within the radius where the temperature
increases above 20~K in a $\rho \propto r^{-1.5}$ envelope around a
5~$L_\odot$ protostar as well as two models where CO sublimates at
30~K and the density profile has either $p=1.5$ with a luminosity of
30~$L_\odot$ or $p=2.0$ and a luminosity of 50~$L_\odot$. These models
are indistinguishable at baselines longer than
15~k$\lambda$. Increasing or decreasing the luminosity by more than
about 20\% makes the fits significantly worse. These models illustrate
some of the degeneracy between the best fit parameters and the
limitations of this method: naturally there is a degeneracy between
the temperature at which CO sublimates and the required luminosity
enhancement -- although this can be somewhat alleviated by considering
the sample in its entirety. Likewise the exact luminosity enhancement
is dependent on the steepness of the density profile: a steeper
density profile causes the C$^{18}$O brightness profile to become
steeper and thereby requires a larger enhancement in luminosity to
shift the sublimation radius outwards. In the following we continue to
consider the case where $p=1.5$ -- but note that the required
luminosity enhancement may be up to a factor 2 larger for steeper density
profiles (i.e., in that context the inferred enhancement is a lower
limit). It is clear that as long as only the longer baselines are
considered the structure and/or abundance on larger scales (shorter
baselines) are unimportant for the fits. This point is further
illustrated in Fig.~\ref{parameterexploration}d where the strength of
the external interstellar radiation field is varied rather than the
internal luminosity of the source: such models may vary the profiles
on the large scales through the sublimation of the CO and changing the
excitation temperature -- but do not affect the shape of the emission
profile in the inner 10$''$ (baselines longer than 15~k$\lambda$) --
corresponding to radii $\lesssim$1000~AU for the studied sources.

In summary, for the test case of IRAS03282 it is not possible to
reproduce the observed C$^{18}$O brightness profiles unless its
abundance profile reflects sublimation at a temperature set by a
higher luminosity of the source than its current $\approx
1$~$L_\odot$.

\begin{table*}[hbt]
\caption{Summary of models from Fig.~\ref{parameterexploration}.}\label{modelsummary}
\begin{center}\begin{tabular}{lccccc}\hline\hline
Panel & Model & Density & \multicolumn{2}{c}{Abundance} & Luminosity \\ \hline
a) & A1 & $\rho \propto r^{-1.5}$ & \multicolumn{2}{c}{$X_0$ (constant)}   &  1.1~$L_\odot$  \\
   & A2 & $\rho \propto r^{-1.5}$ & \multicolumn{2}{c}{$X_0/8$ (constant)} &  1.1~$L_\odot$ \\
   & A3 & $\rho \propto r^{-1.5}$ & \multicolumn{2}{c}{$8\,X_0$ (constant)} & 1.1~$L_\odot$ \\ \hline
b) & B1 & $\rho \propto r^{-1.5}$ & \phantom{$10\,$}$X_0$ for $r\, <\, r_{\rm 20 \,K}$ & $X_0/100$ for $r\, >\, r_{\rm 20 \,K}$ & 1.1~$L_\odot$  \\
   & B2 & $\rho \propto r^{-1.5}$ & \phantom{$10\,$}$X_0$ for $r\, <\, r_{\rm 20 \,K}$ &  $X_0/10$\phantom{$0$}  for $r\, >\, r_{\rm 20 \,K}$ & 1.1~$L_\odot$ \\
   & B3 & $\rho \propto r^{-1.5}$ & $10\,X_0$ for $r\, <\, r_{\rm 20 \,K}$ &  $X_0/10$\phantom{$0$}  for $r\, >\, r_{\rm 20 \,K}$ & 1.1~$L_\odot$ \\
   & B4 & $\rho \propto r^{-1.5}$ & \multicolumn{2}{c}{$X_0$ for $r\,
     <\, r_{\rm 20 \,K}$ -- and  gradual decrease from $X_0$ at $r_{\rm
     out}$ to $X_0/100$ at $r_{\rm 20 \,K}$} & 1.1~$L_\odot$ \\ \hline
c) & C1 & $\rho \propto r^{-1.5}$ & \phantom{$10\,$}$X_0$ for $r\, <\, r_{\rm 20 \,K}$ &  $X_0/10$\phantom{$0$}  for $r\, >\, r_{\rm 20 \,K}$ & 5~$L_\odot$  \\
   & C2 & $\rho \propto r^{-1.5}$ & \phantom{$10\,$}$X_0$ for $r\, <\, r_{\rm 30 \,K}$ &  $X_0/10$\phantom{$0$} for $r\, >\, r_{\rm 30 \,K}$ & 30~$L_\odot$ \\
   & C3 & $\rho \propto r^{-2.0}$ & \phantom{$10\,$}$X_0$ for $r\, <\, r_{\rm 30 \,K}$ &  $X_0/10$\phantom{$0$} for $r\, >\, r_{\rm 30 \,K}$ & 50~$L_\odot$ \\ \hline
d) & D1 & $\rho \propto r^{-1.5}$ & \phantom{$10\,$}$X_0$ for $r\, <\, r_{\rm 20 \,K}$ &  $X_0/100$  for $r\, >\, r_{\rm 20 \,K}$ & 1.1~$L_\odot$  \\
   & D2 & $\rho \propto r^{-1.5}$ & \phantom{$10\,$}$X_0$ for $r\, <\, r_{\rm 20 \,K}$ &  $X_0/100$ for $r\, >\, r_{\rm 20 \,K}$ & 1.1~$L_\odot$ ; ISRF$\times$5\tablefootmark{a}\\
   & D3 & $\rho \propto r^{-1.5}$ & \phantom{$10\,$}$X_0$ for $r\, <\, r_{\rm 20 \,K}$ &  $X_0/100$ for $r\, >\, r_{\rm 20 \,K}$ & 1.1~$L_\odot$ ; ISRF$\times$50\tablefootmark{a} \\ \hline
\end{tabular}\end{center}

\tablefoot{$X_0$ is a C$^{18}$O abundance of $2\times 10^{-7}$ with
respect to H$_2$, corresponding to a canonical CO abundance of
$10^{-4}$. \tablefoottext{a}{Interstellar radiation field (ISRF) enhanced by a factor 5
and 50 above the standard ISRF described by \cite{black94}.}}
\end{table*}

\subsection{Sample approach}\label{approach}
\begin{figure}
\resizebox{\hsize}{!}{\includegraphics{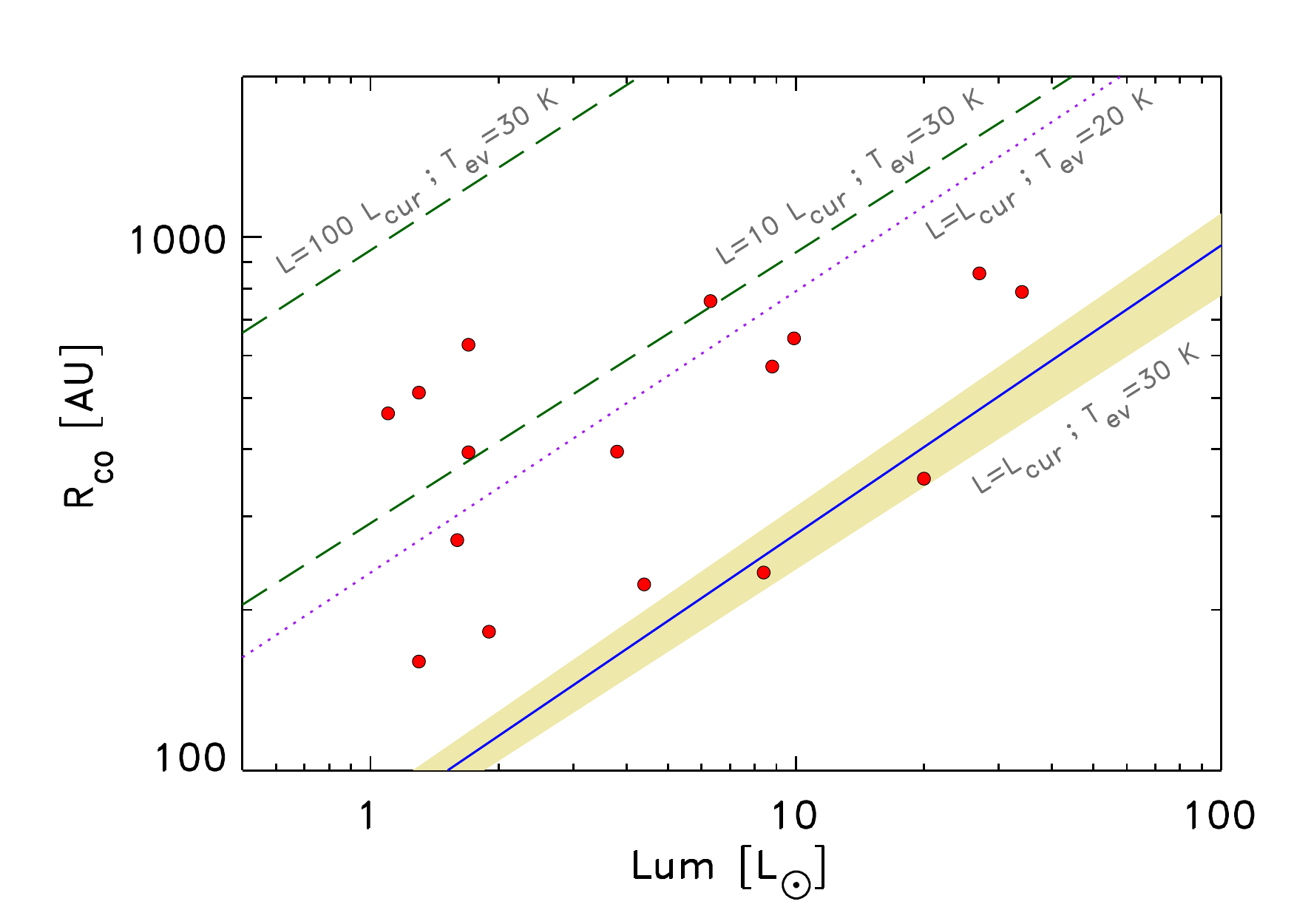}}
\caption{Radii of C$^{18}$O emission at the half-maximum intensity for
  the 16 Class~0 protostellar sources studied in this paper (symbols)
  as function of luminosity compared to the predictions from radiative
  transfer models: the solid line indicates a 1~$M_\odot$ envelope
  with sublimation of CO and resulting increase in abundance by two
  orders of magnitude within the radius where the temperature
  increases above 30~K. The shaded area corresponds to envelope masses
  ranging from 0.3-3~$M_\odot$. The dotted line represents models
  where the sublimation takes place at a temperature of 20~K
  instead. Finally, the dashed lines correspond to sources where the
  luminosity has been increased by one and two orders of magnitude and
  the sublimation takes place for a temperature of 30~K. The
    typical uncertainties on the measured extents are 10--20\% of the
    plotted values; the typical uncertainties in the predicted extents
    are up to 50\% due to the uncertainties in the underlying model,
    luminosity and similar.}\label{summary_figure}
\end{figure}

Inspired by the discussion above we define the approach for the entire
sample of sources. For each source we derive the extent of the
integrated C$^{18}$O 2--1 emission by fitting 2D Gaussian profiles in
the $(u,v)$-plane to limit systematics due to the deconvolution. We
restrict the fit to baselines longer than 15~k$\lambda$ (about 20~m;
or scales corresponding to about 10$''$). As shown above
(\S\ref{testcase}) this selection is ideal for measuring the extent of
the regions of the envelopes where CO sublimates -- and furthermore
filtering the remainder of the low surface brightness extended
emission associated with the ambient cloud. Modeling the emission
  on shorter baselines can be used together with single-dish
  observations to probe the amount of CO in the larger scale envelope
  and, for example, the rate at which CO freezes-out or the importance
  of photo-desorption \citep[e.g.,][]{coevollet,fuente12}. However, as
  we here are mainly interested in the gradients in C$^{18}$O emission
  associated with CO sublimating at higher temperatures and aim to
  treat all sources consistently, we focus only on the emission on the
  longer baselines.  Table~\ref{sampleresults} lists the deconvolved
extent of the C$^{18}$O 2--1 emission (FWHM) for each source and
Fig.~\ref{summary_figure} plots the extents (as the radius =
  FWHM/2) versus the current luminosities of the sources. Already
from this, it can be seen that there is \emph{not} a correlation
between the deconvolved extents and the current luminosities of the
protostars -- what one otherwise would expect if the heating by the
central protostar dominates the temperature profile and thus
determines the size of the region where CO is in the gas-phase.

Fig.~\ref{summary_figure} also shows the predictions for the
C$^{18}$O emission from a set of radiative transfer models. We do
  not construct models for each individual source -- but rather make
  predictions of the C$^{18}$O emission extent from a generic envelope
  model with $\rho \propto r^{-1.5}$ and ranges of luminosities. For
each of these models we follow the same procedure as described in
\S\ref{testcase}, performing the dust and line radiative transfer
calculations and predicting the C$^{18}$O 2--1 emission profile
assuming that the CO comes off dust grains with a two orders of
magnitude change in abundance for a sublimation temperature of 30~K
(see the discussion about a possible lower sublimation temperature
below). The models are processed in the same way described above,
multiplied by the interferometric primary beam field of view,
Fourier-transformed and the resulting extent measured by fitting a 2D
Gaussian-profile in the $(u,v)$-plane. The model predictions are
compared to the observational data in Fig.~\ref{summary_figure} with
Table~\ref{sampleresults} providing the current luminosities, the
predicted sizes of the CO emission profiles given these luminosities,
the actual measured sizes and the corresponding enhancements in
luminosities. The uncertainty of the inferred value for the
  luminosity enhancement from the model fits is about 20\%
  (\S\ref{testcase}). Due to systematic uncertainties (e.g., due to
  the model assumptions such as the envelope mass in the general
  approach), variations should only be considered significant on a
  50\% or higher level -- but generally considered lower limits due to
  the method being insensitive to the most intense bursts due to the
  interferometer resolving out more extended emission and the
  degeneracy with the steepness of the envelope density profiles
  (\S\ref{testcase}).

A few things are immediately apparent from Fig.~\ref{summary_figure}:
the observed extents of the CO emission do not show a clear
correlation with the current luminosities with ``typical'' variations
in the deconvolved sizes for any given luminosity of about a factor of
three. The exception to this statement is that no sources with high
luminosities have very compact C$^{18}$O 2--1 emission. If the extent
of the C$^{18}$O emission is determined primarily by the sublimation
due to heating by the central protostar this is unsurprising, however:
as the sublimation is expected to be instantaneous compared to the
evolutionary time-scales of the protostars -- it is unlikely that one
encounters such sources.  It is indeed seen that the sources with the
most compact emission for each luminosity nicely track the predictions
from the radiative transfer models. In contrast, for about half the
sources the CO emission is seen to be extended as one would expect if
the luminosity had been from a factor about five up to two orders of
magnitude higher than it is now. Under the same interpretation this is
an indication that their current luminosities are not what determines
the observed extent of the CO emission.

\begin{table*}
\caption{Measured and predicted extents of the deconvolved C$^{18}$O
  emission -- as well as the current luminosities and luminosity
  enhancements required to reproduce the extended C$^{18}$O emission.}\label{sampleresults}\centering
\begin{tabular}{lllll}\hline\hline
Source    & $L_{\rm cur}$ ($L_\odot$) & Predicted size ($''$)\tablefootmark{a} &
Measured size ($''$)\tablefootmark{b} & $L_{\rm CO}/L_{\rm cur}$\tablefootmark{c} \\ \hline
L1448I2   &  1.7 &  0.9 &  5.3 &   25 \\
L1448I3   &  8.8 &  2.2 &  4.9 &  4.2 \\
L1448C    &  8.4 &  2.2 &  2.0 &  0.8 \\
IRAS03256 &  1.7 &  0.9 &  3.4 &   12 \\
IRAS2A    &   20 &  3.3 &  3.0 &  0.8 \\    
SVS13A    &   34 &  4.6 &  6.7 &  2.0 \\    
IRAS4A    &  9.9 &  2.4 &  5.5 &  4.6 \\    
IRAS4B    &  4.4 &  1.6 &  1.9 &  1.4 \\    
IRAS03282 &  1.1 &  0.8 &  4.0 &   25 \\    
TMR1      &  3.8 &  2.4 &  5.7 &  5.4 \\    
L1527     &  1.9 &  1.6 &  2.6 &  2.3 \\       
MMS6      &   27 &  2.0 &  4.1 &  3.7 \\       
MMS9      &  6.3 &  1.1 &  3.6 &  9.6 \\       
IRAS15398 &  1.6 &  1.3 &  3.5 &  5.9 \\       
CrA32     &  1.3 &  1.5 &  7.9 &   24 \\       
B335      &  1.3 &  1.9 &  3.2 &  2.7 \\ \hline
\end{tabular}

\tablefoot{\tablefoottext{a}{Predicted extent of the C$^{18}$O 2--1 emission given the
    current luminosity of the source ($L_{\rm cur}$).}
  \tablefoottext{b}{Measured
    deconvolved extent of the C$^{18}$O 2--1 emission (FWHM). The uncertainties are dictated by the
    signal-to-noise and the assumption of the emission being
    distributed as Gaussians: typical statistical uncertainties are
    0.1--0.5$''$ -- or 10--20\% of the measured sizes.}
  \tablefoottext{c}{Ratio of luminosity needed to reproduce
    C$^{18}$O 2--1 extent to current luminosity -- i.e., inferred
    enhancement in luminosity. See the text for discussion about the assocated
    uncertainties.}}
\end{table*}

\section{Discussion}\label{discussion}
\subsection{Envelope mass and CO sublimation temperature}\label{caveats}\label{sublimationtemperature}
Two parameters that may affect the exact value of the inferred
increase in luminosity on the basis of the CO extent are the assumed
envelope mass and CO sublimation temperature in the generic models.

A simple test reveals that the former is of little importance for the
solar-type protostars considered here: Fig.~\ref{summary_figure} shows
the spread in inferred CO sublimation radii for envelope masses
varying by an order of magnitude from 0.3 to 3~$M_\odot$: the
variation in the predicted CO sublimation radii is about 10\%. This
reflects that the envelopes are largely optically thin to their own
radiation in which case the temperature profile can be described by a
power-law temperature profile with radius with the overall scaling
only determined by the luminosity of the source
\citep[e.g.,][]{scoville76,doty94,chandler00}.

The adopted CO sublimation temperature is much more important as also
illustrated in Fig.~\ref{summary_figure}. Shifting the sublimation
temperature from 30~K to 20~K in the models shifts the radius where CO
evaporates out to align with the predictions for where it would be for
a model with a sublimation temperature of 30~K but luminosities an
order of magnitude higher.

This discussion is highly relevant: traditionally many studies of
modeling sublimation of CO in star-forming regions adopt a binding
energy of 855--960~K from laboratory experiments of pure CO
\citep[e.g.,][]{sandford93,bisschop06}. However, laboratory
experiments also show that mixing the CO with H$_2$O ice can increase
the CO binding energy to $\approx$~1200--1300~K
\citep[e.g.,][]{collings03,noble12}. The thermal sublimation rate at a
constant dust temperature, $T_d$, depends exponentially on the binding
energy, $E_b$, as $\xi\propto \exp(E_b/T_d)$ (e.g., Eq.~4 of
\citealt{rodgers03}). For a binding energy of 960~K it takes less than
1~year for CO to sublimate for a dust temperture above 21~K. For a
binding energy of 1300~K the sublimation rates slows down so that it
takes less than 1~year for CO to sublimate only at temperatures above
28~K. Observationally, there is some evidence for a higher sublimation
temperature based on modeling of multi-transition CO observations as
well as previous high angular resolution observations
\citep{l483art,coevollet,yildiz13}.

It is worth noting that half of the sources in fact do show compact CO
emission that is well-fit by the models with a sublimation temperature
of 30~K adopting their current source luminosities. That directly
argues against a lower sublimation temperature as it would result in
more extended CO emission than observed. In contrast, a higher
sublimation temperature cannot be ruled-out based on the observations
here -- but it would imply a larger fraction of the sources have
experienced increases in luminosities in their recent histories. 

This result is not necessarily a contradiction with some high
resolution imaging studies of more evolved disks around T~Tauri stars
that may favor lower sublimation temperatures through indirect imaging
\citep[e.g.,][]{qi13,mathews13} -- although it might be interesting to
revisit those studies with models for which a higher binding
energy/sublimation temperature would be adopted. Whereas the CO and
H$_2$O ices are likely formed simultaneously in the protostellar
environments where the temperatures generally are low -- the ices in
the disks around T Tauri stars may have undergone some heating and
sublimation before reformation when entering the disks and the
material cooling down. As the binding energy of H$_2$O is much higher
than that of CO, only a small fraction of the water may not sublimate
before entering the disk \citep[e.g.,][]{visser09,cleeves14}. This
would naturally lead to the formation of a clearer onion-shell
structure of the ices in the disks with an inner layer of H$_2$O ice
and an outer layer of more purified CO ice with a resulting lower
binding energy. It is possible that the CO sublimation
  temperature varies from source-to-source, e.g., depending on the
  water content in their ambient environment. This would naturally be
  an important result in itself -- although, no other direct
  observational evidence exists for such variations. In that case, one
  could only argue that three out of the sixteen sources would show
  evidence for enhanced luminosities in their recent histories.

In any case, this discussion emphasizes the need for high quality
laboratory measurements of ices relevant for astrophysical
environments -- and a careful consideration of the exact conditions
and chemical histories of the studied regions.

\subsection{Ice sublimation as a tracer of accretion bursts}
\begin{figure*}
\sidecaption
\includegraphics[width=12cm]{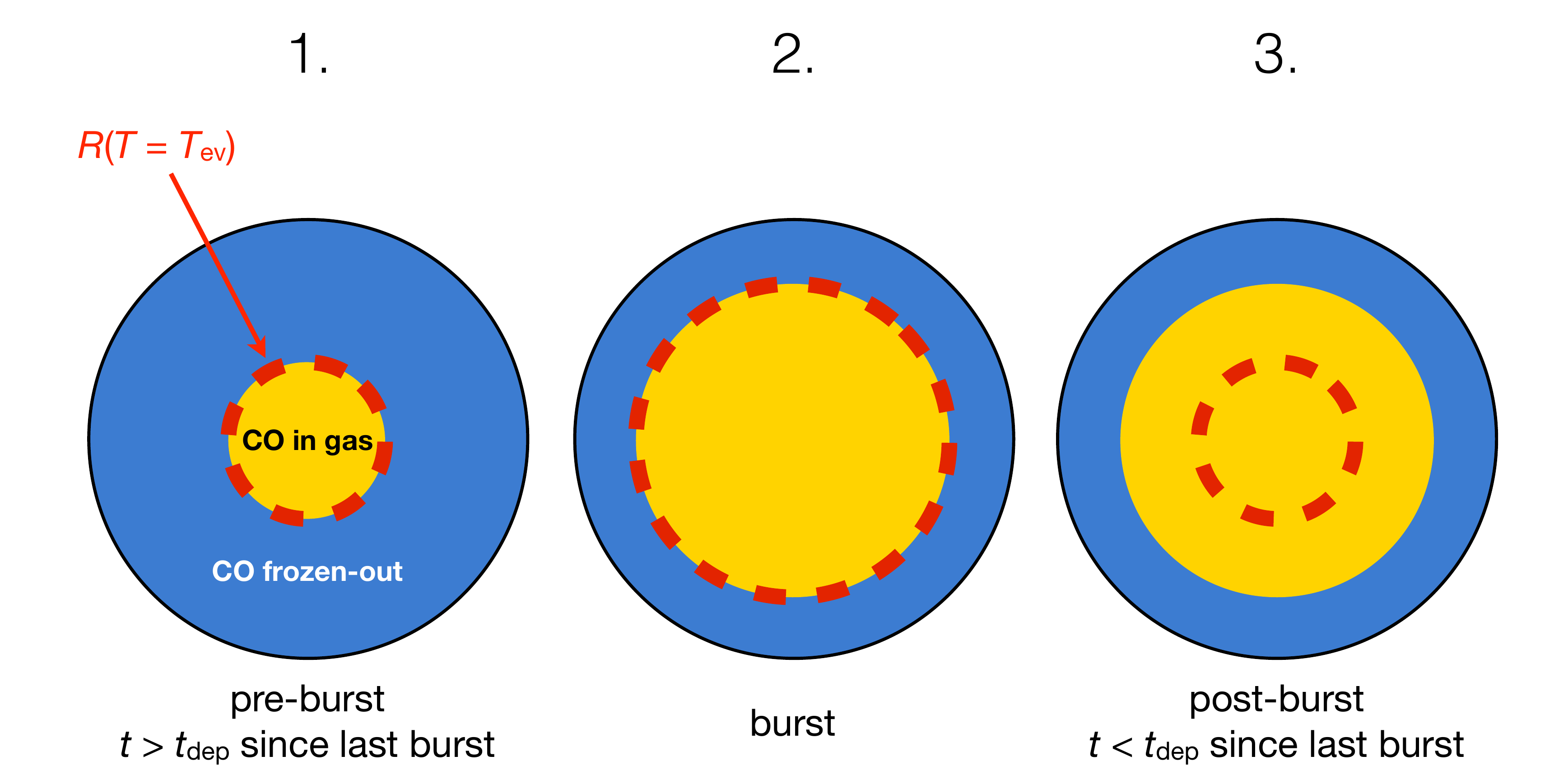}
\caption{Cartoon illustrating the interpretation of the extended CO
  emission signatures due to variations in source luminosities.  The
  dark blue and yellow regions indicate where CO is frozen-out and in
  the gas-phase in three characteristic situations around a
  protostellar burst in luminosity. The red dashed line is the radius
  where the temperature in the envelope reaches the sublimation
  temperature for CO. The three sketches indicate the situations (1)
  just before a burst where the CO sublimates out to a radius defined
  by the current luminosity of the protostar, (2) during a burst where
  the luminosity of the protostar increases and the CO sublimation
  radius moves outwards, and (3) immediately after a burst where CO
  remains in the gas-phase in a large region -- before it freezes out
  again.}\label{cartoon}
\end{figure*}

A fairly straightforward interpretation of the variations in the
extent of the central C$^{18}$O emission relative to the predictions
given the current source luminosities is that the protostars in fact
undergo bursts of accretion and corresponding
luminosities. Fig.~\ref{cartoon} illustrates this progression through
three characteristic phases: in a period of time before the burst, or
equivalently a long time since the previous burst, the chemistry will
set itself in a relatively steady state with CO frozen-out outside the
radius where the temperature reaches the sublimation temperature and
in the gas-phase within this (Stage~1 in Fig.~\ref{cartoon}). During
the burst this radius will be shifted outwards and due to the short
time-scale for sublimation of CO the freeze-out/sublimation boundary
will shift outwards correspondingly (Stage~2). For typical binding
energies for CO (see \S\ref{sublimationtemperature}) the time-scale
for CO to sublimate at temperatures of 30~K is of order minutes to
days -- i.e., much shorter than the typical dynamical time-scales
related to the infall of material from the envelope. Finally, after
the burst has finished and the source reset to a low luminosity, CO
will begin to freeze-out again. However, due to the slower freeze-out
it will still appear to be extended for a longer period of time
(Stage~3). 

It is possible to be more quantitative about the time-scales. For a
characteristic envelope with a mass of 0.3--0.5~$M_\odot$, a size of
10,000~AU (radius) and a density profile dropping as $\rho \propto
r^{-1.5}$, the time-scale for CO to freeze-out at the density at
300--500~AU is of order $t_{\rm dep}\sim 10^4$~years (e.g., Eq.~3 of
\citealt{rodgers03}). Thus, if a given source undergoes an increase in
luminosity, CO would stay in the gas-phase -- and its emission appear
extended -- for this period of time after the luminosity has decreased
again. 

With additional assumptions one can even turn this into a statement
about the frequencies of such bursts implied by the current data. One
can for example assume that the duration of the burst in Stage~2
($t_2$) from Fig.~\ref{cartoon} is much shorter than the total
duration of Stage~1 through 3 ($t_1+t_2+t_3$): based on the observed
luminosity distributions of young stellar objects from the Spitzer
Space Telescope c2d legacy program and simple calculations of the
accretion rates, \cite{evans09} argued that half the final mass of a
given young star is accreted during 7\% of the duration of the
embedded stages. Similarly from a comparison between numerical
hydrodynamical simulations of collapsing cores and the distributions
populations of young stars, \cite{dunham12} argued that modeled
protostars on average spend $\approx$1\% of their lifetimes in modes
of accretion bursts.  The analysis above shows that half of the
sources have CO extents corresponding to luminosities about a factor
five or more above the current. In context of the scenario outlined
above this would imply that $t_1 \approx t_3$. As the duration of
$t_3$ per definition is the depletion time, $t_{\rm dep}$, this
suggest that a given source undergoes a burst every $t_1+t_3 \sim 2
t_{\rm dep} \approx$~20,000~years -- or of order 5 bursts during
the $10^5$~year duration of the deeply embedded Class~0 stage. If
  one on the other hand assumes that variations in the CO sublimation
  temperatures between the different sources is the dominant reason
  for the variations in the CO extent and thus that only three out of
  the sixteen sources show significant C$^{18}$O 2--1 extents, this would
  imply a burst every 50,000~years.  These time-scales are in
good agreement with estimates by \cite{scholz13} who compared Spitzer and
WISE photometric data for young stars in nearby star forming regions
and identified 1--4 bursts in a sample of 4000 young stars.
Translated into frequencies these numbsers would correspond to bursts
in intervals of about 5000 and 50,000~yr.

It is worth re-emphasizing the reasons why the method outlined in this
paper is useful: first and foremost the C$^{18}$O emission is not
significantly affected by, e.g., outflows or the ambient
environment. This is observationally shown given the narrow lines and
relatively centrally condensed distributions of the line
emission. This is mainly a result of the C$^{18}$O being largely
optically thin and thus predominantly sensitive to the high column
densities of material associated with the protostars themselves. An
analysis of the chemistry of common molecular tracers during and after
the accretion burst is presented by \cite{visser15}. Also, any more
smoothly distributed material on larger scales in the clouds is
filtered by the interferometer.  Secondly, the time-scales for
sublimation and freeze-out of the CO molecule are apparently
  well-matched for its use as a tracer. Observationally the typical
  envelope masses (and consequently densities) change on time-scales
  of a few $10^5$~years from one to a few $M_\odot$ to 0.1 to
  0.5~$M_\odot$ for early Class I sources, which is still sufficiently
  long for freeze-out and sublimation to occur. In fact, the
observation that some sources do show the compact C$^{18}$O emission
directly shows that large variations on short time-scales are unlikely:
had the frequency of bursts been higher than $1/t_{\rm dep}$ the
sources would never get back to the Stage~1 and cycle between Stages~2
and 3 -- with the result that the CO emission would always appear
extended.

Of course these estimates are still rather crude due to the low number
sample statistics. Also, due to the inherent uncertainties in the
modeling discussed above, the analysis does not constrain the
strengths of the bursts. For example, due to the low sublimation
temperatures the CO isotopologues are not sensitive to very intense
bursts that may cause it to evaporate out to the largest scales probed
by the interferometric data ($\sim 1000$~AU radius). Observations of
other species that sublimate at higher temperatures (or species
reacting to those) could help refine the accretion histories of
protostars. Observations of the H$_2$O could for example be used
  to directly test whether the binding energy of CO varies
  significantly from source to source: if the extents of the H$_2$O
  emission is found to vary in conjunction with CO, it would indicate
  that luminosity variations dominate over variations in the binding
  energy. Surveys of larger (unbiased) samples of sources coupled to
models would greatly improve the statistics also on smaller scales
and/or correlate to, e.g., physical properties of the ambient
environment or the physical structure of the protostars on
small-scales that may drive possible accretion variations. A more
  detailed description of the dynamical evolution of the protostellar
  envelopes on the characteristic $10^4$~year time-scale for
  freeze-out would also be needed to refine the statistics. 
  Finally, although the maps do not show any indications that the
  C$^{18}$O 2--1 emission is affected by the presence of outflows,
  sensitive high angular resolution observations will be needed to
  quantify what impact outflows have on the quiescent gas and how they
  contribute to shaping the envelopes.

Turning the argument around: results such as these emphasize the need
to take the accretion histories of embedded protostars into account
when considering their chemistry. If similar accretion variations can
be confirmed to take place during the evolution of embedded protostars
in general they may have profound implications for the gas-grain
chemistry interplay taking place during the early stages of
protostellar evolution.

\section{Summary}\label{summary}
An analysis of maps of C$^{18}$O $J=$~2--1 emission from the
Submillimeter Array (SMA) toward a sample of deeply embedded
protostars has been presented. The maps trace the distribution of
C$^{18}$O in the envelope on few hundred to thousand AU scales. The
maps are compared to detailed continuum and line radiation transfer
models -- in particular, the predictions for the extent of the CO
emission given the current luminosities of the sources. The
conclusions are as follows:
\begin{itemize}
\item The C$^{18}$O spectra appear relatively symmetric with little
  evidence for strong outflow action. The integrated emission maps are
  likewise regular with only faint low surface brightness extended
  emission. These results indicate that the integrated C$^{18}$O
  emission is predominantly picking up the material in the denser
  parts of the envelopes on $\lesssim 1000$~AU scales.
\item A case study of the IRAS03282 protostar is used to demonstrate a
  method in which the deconvolved extent of the C$^{18}$O emission
  measured in the $(u,v)$-plane can be compared to radiation transfer
  models for the protostellar envelope. The key parameter in this
  comparison is the assumed luminosity of the source and consequently
  the radius at which CO sublimates at temperatures of 20--30~K. This
  dominates over other effects such as the envelope physical structure
  and exact absolute abundances.
\item In the comparison for the entire sample it is found that half of
  the sources show C$^{18}$O line emission that is more extended than
  predicted by models based on the current luminosities of the sources
  if CO sublimates at temperatures of 30~K. For those sources the
  increase has to have been more than about a factor five; for four of
  these the inferred luminosity would be more than a factor 10 higher.
\item A CO sublimation temperature lower than 30~K cannot be
  reconciled with the observations of half the sources with compact
  emission. The sublimation temperature of 30~K in these environments
  may be taken as evidence of a higher binding energy for the CO than
  what is expected for pure ices -- a possibility if CO is mixed in
  with CO$_2$ and/or H$_2$O.
\item The half of the sources in the sample that show more extended
  C$^{18}$O emission are the candidates for those that have undergone
  a recent change in luminosity, e.g., due to a burst in accretion
  rate. This would have had to take place within the last $\sim
  10^4$~years, which is the time it takes for CO to freeze-out on dust
  grains at the densities characteristic of the 300-500~AU radii of
  the protostellar envelopes observed here. If such bursts of
  accretion are taking place in general, the small number statistics
  based on this sample would imply that a protostar undergoes of order
  5 bursts during its first $10^5$~years. If one assumes that the
    CO sublimation temperature varies between 20 and 30~K from
    source-to-source, a conservative estimate would be an interval of
    50,000~years between the bursts.
\end{itemize}

This study highlights the importance of studying the physical and
chemical evolution of embedded protostars in unison. CO is one of the
key species in the gas-phase and grain-surface chemistry taking place
in the environments of these protostars -- and understanding the
physical mechanisms leading to its freeze-out and sublimation
therefore of key importance and something that needs to be taking into
account any models for protostellar chemistry. Conversely, a good
understanding of its chemistry may be used as a diagnostic tool for
effects such as variations in the accretion histories of the
protostars. An important next step would be to extend the current
  study to a more systematically selected sample of sources and
  differing molecular species, e.g., probing different temperature and
  density regimes, to for example break the degeneracies introduced by
  the limitations in our understanding of the details of the chemistry
  of just one species.  In the near-future ALMA will make this
possible and thereby open up the potential for investigating the
``dynamic chemistry'' in star and planet-forming regions. In this
context it will provide excellent maps for a large number of sources
and species and provide good statistics also revealing the magnitudes
of possible accretion bursts.

\begin{acknowledgements}
  We thank the referee for a thorough report that greatly helped the
  presentation and discussion of the results. This paper is based on
  data from the Submillimeter Array: the Submillimeter Array is a
  joint project between the Smithsonian Astrophysical Observatory and
  the Academia Sinica Institute of Astronomy and Astrophysics and is
  funded by the Smithsonian Institution and the Academia Sinica. The
  research of JKJ was supported by a Junior Group Leader Fellowship
  from the Lundbeck foundation. Research at Centre for Star and Planet
  Formation is funded by the Danish National Research Foundation and
  the University of Copenhagen's programme of excellence.
\end{acknowledgements}

\end{document}